\pdfoutput=1
\documentclass[journal=jacsat,manuscript=article]{achemso}

\usepackage[version=3]{mhchem} 
\usepackage{upgreek}
\usepackage{textcomp} 
\usepackage{cleveref}
\usepackage{xcolor}
\usepackage{ulem}
\usepackage{comment}

\author{Marion J. L. Sourribes}
\affiliation
{London Centre for Nanotechnology, University College London, London WC1H 0AH, United Kingdom}
\author{Ivan Isakov}
\affiliation
{London Centre for Nanotechnology, University College London, London WC1H 0AH, United Kingdom}
\author{Marina Panfilova}
\affiliation
{London Centre for Nanotechnology, University College London, London WC1H 0AH, United Kingdom}
\author{Huiyun Liu}
\affiliation{Department of Electronic and Electrical Engineering, University College London, London WC1E 7JE, United Kingdom}
\author{Paul A. Warburton}
\affiliation{London Centre for Nanotechnology, University College London, London WC1H 0AH, United Kingdom}
\alsoaffiliation{Department of Electronic and Electrical Engineering, University College London, London WC1E 7JE, United Kingdom}
\email{p.warburton@ucl.ac.uk}

\title
  {Mobility Enhancement by Sb-mediated Minimisation of Stacking Fault Density in InAs Nanowires Grown on Silicon}


\begin{document}
\begin{abstract}

We report the growth of InAs$_{1-x}$Sb$_{x}$ nanowires (0\,$\leq$\,x\,$\leq$\,0.15) grown by catalyst-free molecular beam epitaxy on silicon (111) substrates. 
We observed a sharp decrease of stacking fault density in the InAs$_{1-x}$Sb$_{x}$ nanowire crystal structure with increasing antimony content. 
This decrease leads to a significant increase in the field-effect mobility, this being more than three times greater at room temperature for InAs$_{0.85}$Sb$_{0.15}$ nanowires than InAs nanowires.

\end{abstract}

\clearpage


Semiconductor nanowires are leading candidates for future applications in a wide variety of electronic, photonic and sensing devices~\cite{Lieber_NanoLett01, Samuelson_NanoLett01}.
III-V compound semiconductor nanowires including  InAs~\cite{Abay_NanoLett01} and GaN~\cite{FET6, LED3} have a number of potential functional advantages over elemental semiconductor nanowires including high mobility and direct bandgap. Furthermore the magnitude of the bandgap can be modulated by exploiting ternary compound semiconductors (such as InAsP~\cite{IVAN-PAPER} and AlGaAs~\cite{AlGaAs_01}), allowing the creation of heterostructure nanowires with axially- or radially-modulated electronic properties. Such bandgap engineering is in principle a more powerful tool for nanowire-based devices than thin-film-based devices since the radial relaxation of strain in nanowires allows the growth of heterostructures whose constituent compounds are significantly lattice-mismatched~\cite{LatticeMismatchNW02,Glas1}.

The growth of compound semiconductor nanowires directly onto single crystal silicon wafers would be advantageous,~\cite{SelectiveArea8, MixedStructureInInAs01} because (i) it would allow integration of nanowire devices with the established silicon CMOS technology; and (ii) silicon wafers are orders of magnitude cheaper than their compound semiconductor counterparts. Compound semiconductor nanowires are, however, typically grown using the ``vapor-liquid-solid'' technique in which gold nanoparticle catalysts seed the growth. Gold cannot be combined with silicon since it forms trap states in the silicon bandgap~\cite{PhysRevLett.44.606, GoldInSilicon02, GoldInSilicon03}. 
Nickel has also been used to catalyze InAs nanowire growth on silicon~\cite{InAsNWOxideLayerThickness01} but these nanowires are not functional for direct integration as they grow following random orientations with respect to the substrate. There have therefore been many reports of nanowire growth without the use of heterocatalytic nanoparticle seeds~\cite{MandlGrowth, MixedStructureInInAs02, SelfGrowth2, SelfGrowth7, SelfGrowth6-Huyiun, InAsWithCatalyst_01} . In the case of the widely-studied narrow-bandgap semiconductor InAs, however, the absence of a heterocatalyst results in the nanowires displaying very high densities of defects including stacking faults, twin boundaries and polytypism, i.e. uncontrolled axial modulation of the crystal structure between the zinc-blende (cubic) and the wurtzite (hexagonal) polytypes of InAs~\cite{InAsWithCatalyst_01, MixedStructureInInAs02}. This in turn leads to an undesirable suppression of the electron mobility~\cite{PhaseMixingInAsNW}.

In this letter, we investigate an approach to reduce the defect density in catalyst-free InAs nanowires via the incorporation of antimony during the growth. We report for the first time the growth of catalyst-free InAs$_{1-x}$Sb$_{x}$ nanowires (0\,$\leq$\,x\,$\leq$\,0.15) on silicon. Due to their tunable narrow band-gap in the mid-infrared spectrum, InAs$_{1-x}$Sb$_{x}$ nanowires grown on Si make prime candidates for the fabrication of highly competitive and eco-friendly infrared photodetectors~\cite{ReviewInfraredDetector}. We quantitatively determine the variations in crystal structure and defect density depending on the antimony incorporation before studying their effect on the nanowire electrical properties.
The antimony incorporation suppresses the hexagonal phase and reduces the stacking fault density by up to one order of magnitude. The reduction of the stacking-fault density results in a large increase in the electron mobility above that of pure InAs nanowires.
While an enhancement in the mobility of Au-nucleated InAs$_{1-x}$Sb$_{x}$ nanowires compared to InAs nanowires has previously been observed~\cite{thelander:232105}, this letter is the first report of the combination of advanced structural and electrical characterizations in \textit{gold-free} InAs nanowires.
By advancing crystal phase control in gold-free nanowires, our work will promote the development of future InAs nanowire-based devices directly integrated with silicon CMOS circuits.

\section{Experimental Details}
\vspace{0.3cm}

All InAs$_{1-x}$Sb$_{x}$ nanowires were grown on p-type Si (111) substrates without the use of catalysts in a Veeco molecular beam epitaxy (MBE) system equipped with a solid In source and As$_4$ and Sb$_2$ cracker cells. 
After an initial annealing of the substrate for 8\,min at 760\,\textdegree C under a constant arsenic flux with a beam equivalent pressure (BEP) of 0.8\,--\,1.0$\times 10^{-5}$\,Torr, the temperature is lowered to 450\,--\,480\,\textdegree C. 
Indium is introduced into the chamber with a BEP of $4.3 \times 10^{-8}$\,Torr for 10\,min to form 150\,--\,200\,nm long InAs nucleation nanowires. The antimony supply is then additionally opened to start the InAs$_{1-x}$Sb$_{x}$ growth with the Sb$_2$ BEP ranging from  0.3 to 1.0$\times 10^{-7}$\,Torr.
After 110\,min the growth is terminated by closing the indium and antimony supplies before cooling down the sample under an arsenic flux. For reference, InAs nanowires were similarly grown via a catalyst-free process on Si (111) substrates. 
\ref{tab:sumGrowthShort} summarizes the growth parameters for different nanowire samples (more details available in Supporting Information).
\begin{table}
  \caption{Growth parameters and morphology of InAs$_{1-x}$Sb$_{x}$ nanowires}
  \label{tab:sumGrowthShort}
  \begin{tabular}{cccccc}
    \hline
     & T  &  FF$_\text{Sb}$\textsuperscript{\emph{a}}  &    diameter & length & Sb content \\
     & (\textdegree C)                                 &  (\%) &  (nm) & ($\upmu$m) & (\%) \\ 
    \hline
InAs & 450                & 0  &  60-125 & 1.8--4 & 0 \\
\hline
 & 480  & 0.40  & 60--180 & 1.7--2.5 & 3.9\\
 & 480  &  0.72 & 70--120 &  1.3--2.4 & 6.6\\
InAs$_{1-x}$Sb$_{x}$ & 480  &  0.98 & 70--115 &  1.4--3.0 & 7.5\\
 & 480   & 1.31 & 100--210 & 2.2--4.3 & 7.9 \\
 & 480  & 1.53 & 130--200 & 2.0--4.5 & 7.7 \\
 & 450  & 1.53  & 80--130 & 1.0--2.3 & 15.4 \\
    \hline
  \end{tabular}

  \textsuperscript{\emph{a}} Antimony fractional flux
\end{table}
The incorporation of antimony in the nanowires was controlled by varying the antimony fractional flux (FF$_\text{Sb}$) representing the ratio of Sb flux to the combined (As and Sb) group-V material flux:
$ \text{FF}_\text{Sb} =F(\text{Sb$_2$})/[F(\text{Sb$_2$}) + F(\text{As$_4$})]$ where \textit{F}(Sb$_2$) and \textit{F}(As$_4$) are respectively the fluxes of antimony and arsenic.
The Sb-content was extracted from high-resolution X-ray diffraction (HRXRD) measurements.
\begin{figure}[ht]
\includegraphics[width=0.5\textwidth]{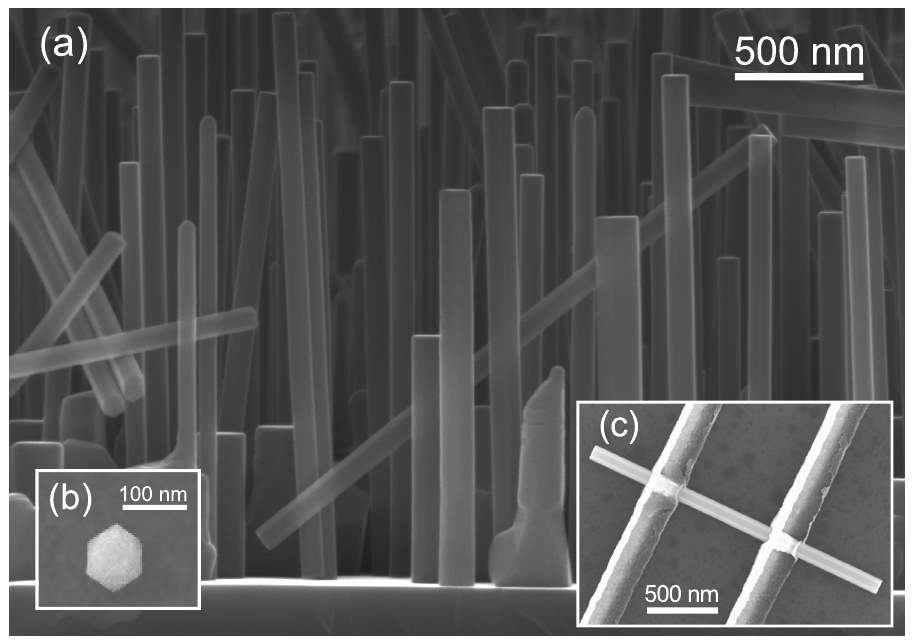}
\caption{(a) SEM image of InAs$_{0.85}$Sb$_{0.15}$ nanowires grown by MBE on a silicon (111) substrate. (b) Top view of a nanowire. (c) SEM image of an InAs$_{0.85}$Sb$_{0.15}$ nanowire field-effect transistor. }
\label{fig:nwForest}
\end{figure}
\ref{fig:nwForest}(a) shows an SEM image of InAs$_{0.85}$Sb$_{0.15}$ nanowires on a Si substrate after growth. The nanowires are generally vertically oriented on the substrate and have regular hexagonal cross-sections, as shown in~\ref{fig:nwForest}(b). Their length varies from 1.2 to 4.5\,$\upmu$m. The diameter remains constant along the length of the wire and varies between 60 and 210\,nm from one nanowire to another.
After growth the nanowires are dispersed in isopropyl alcohol and transferred onto degenerately B-doped silicon substrates coated by a thermally grown 250\,nm thick silicon dioxide layer with pre-patterned macroscopic metal pads. An electron-beam sensitive resist (PMMA 950KA4, 300\,nm thick layer) is spin-coated on the sample. Electron-beam lithography is then used to define contacts in two-point configuration between the selected nanowires and the existing pads. The contacts have a width of 250\,nm and are separated by between 0.75 and 1.1\,$\upmu$m. To remove the native oxide surrounding the nanowires and assure highly transparent contacts, the contact area of the nanowires was treated by argon milling with a dose of 0.21\,C.cm$^{-2}$~\cite{MY-PAPER-ContactResistance}. Only the contact area is treated during the argon milling process, as the remainder of the nanowire is protected by the PMMA resist.
The argon milling is directly followed by an in situ sputter deposition of a 100\,nm thick niobium layer. 
\ref{fig:nwForest}(c) shows a typical InAs$_{0.85}$Sb$_{0.15}$ device nanowire after connection of the contacts.

\section{Sb incorporation and structural characterization of InAs$_{1-x}$Sb$_{x}$ nanowires}
\vspace{0.3cm}

\begin{figure}[ht!]
\includegraphics[]{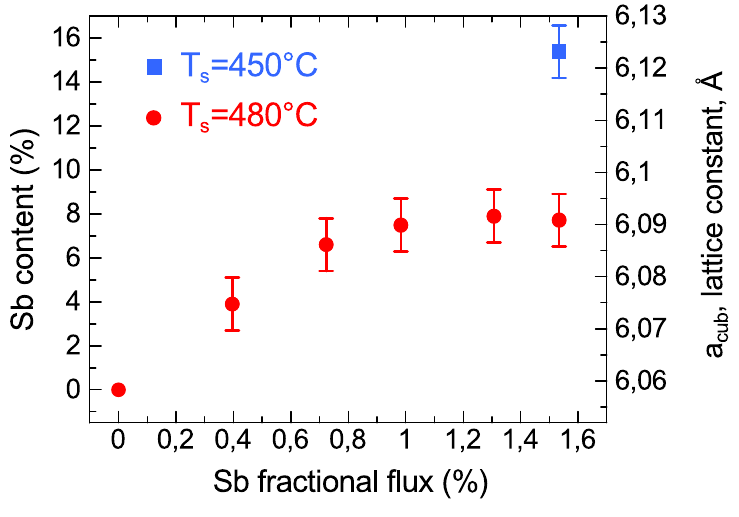}
\caption{Antimony content and lattice parameter in InAs$_{1-x}$Sb$_{x}$ nanowires as a function of the Sb fractional flux for two growth temperatures (blue square for 450\,\textdegree C and red circles for 480\,\textdegree C).}
\label{fig:MarinaXX}
\end{figure}

\begin{figure}[ht!]
\includegraphics[]{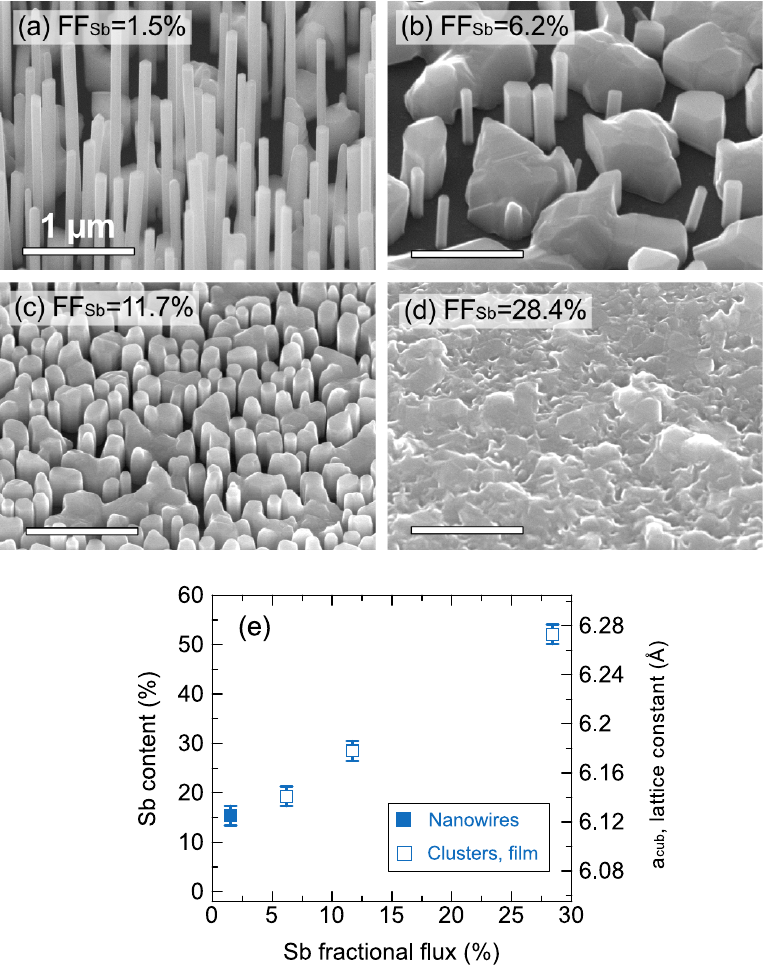}
\caption{(a--d) SEM images of InAs$_{1-x}$Sb$_{x}$ samples grown at 450\,\textdegree C with various Sb fractional flux. The scale bar is 1\,$\upmu$m for all images. (e) Antimony content and lattice parameter in InAs$_{1-x}$Sb$_{x}$ samples grown at 450\textdegree C as a function of the Sb fractional flux. }
\label{fig:450Growths}
\end{figure}

High resolution XRD measurements were used to study the solid composition of the InAs$_{1-x}$Sb$_{x}$ nanowires. From the angular position of the InAs$_{1-x}$Sb$_{x}$ peak, the lattice parameter was calculated by using Bragg's law and the antimony content was subsequently extracted by using Vegard's law~\cite{VegardsLaw,IVAN-PAPER}. 
\ref{fig:MarinaXX} shows the extracted lattice parameters and antimony content in InAs$_{1-x}$Sb$_{x}$ nanowires as a function of the Sb fractional flux for different growth temperatures. We estimate an overall uncertainty of less than 0.01\,\AA{} in the estimation of the change in lattice constant accounting for sample misalignment, detector and goniometer resolution and uncertainty in the position of the Bragg peak due to the presence of the short InAs stems. The influence of the two-dimensional layer grown along the nanowires was investigated by removing the nanowires from the samples with sticky tape. The peak intensity obtained for the stripped samples was almost two orders of magnitude lower than the peak obtained in presence of nanowires, therefore allowing us to neglect the presence of the two-dimensional layer for the determination of the nanowire lattice constant. For samples grown at 480\,\textdegree C, the nanowire lattice parameter increases when increasing the Sb fractional flux from 0 to 1.3\%. Above 1.3\%, the antimony incorporation saturates at around 8.0\%. By decreasing the growth temperature to 450\,\textdegree C, a 15.4\% incorporation was achieved.
Varying the Sb fractional flux from 1.5\% to 28.4\% at 450\,\textdegree C led to a significant increase in the antimony incorporation, up to 52\%, as seen in \ref{fig:450Growths}. However, nanowires become much shorter ($\sim$0.3--0.7\,$\upmu$m) for higher fractional flux. Clusters develop on the substrate until the formation of a bare InAs$_{0.48}$Sb$_{0.52}$ film. Antimony atoms diffuse more slowly at 450\textdegree C than at 480\textdegree C which could favor the development of clusters to the detriment of nanowire growth. Surface preparation of the growth substrate with the patterning of holes in a SiO$_2$ mask, as commonly used in selective area epitaxy~\cite{SelectiveArea2}, could be a possible solution to this issue.

\vspace{0.5cm}

\begin{figure}[ht!]
\includegraphics[width=\textwidth]{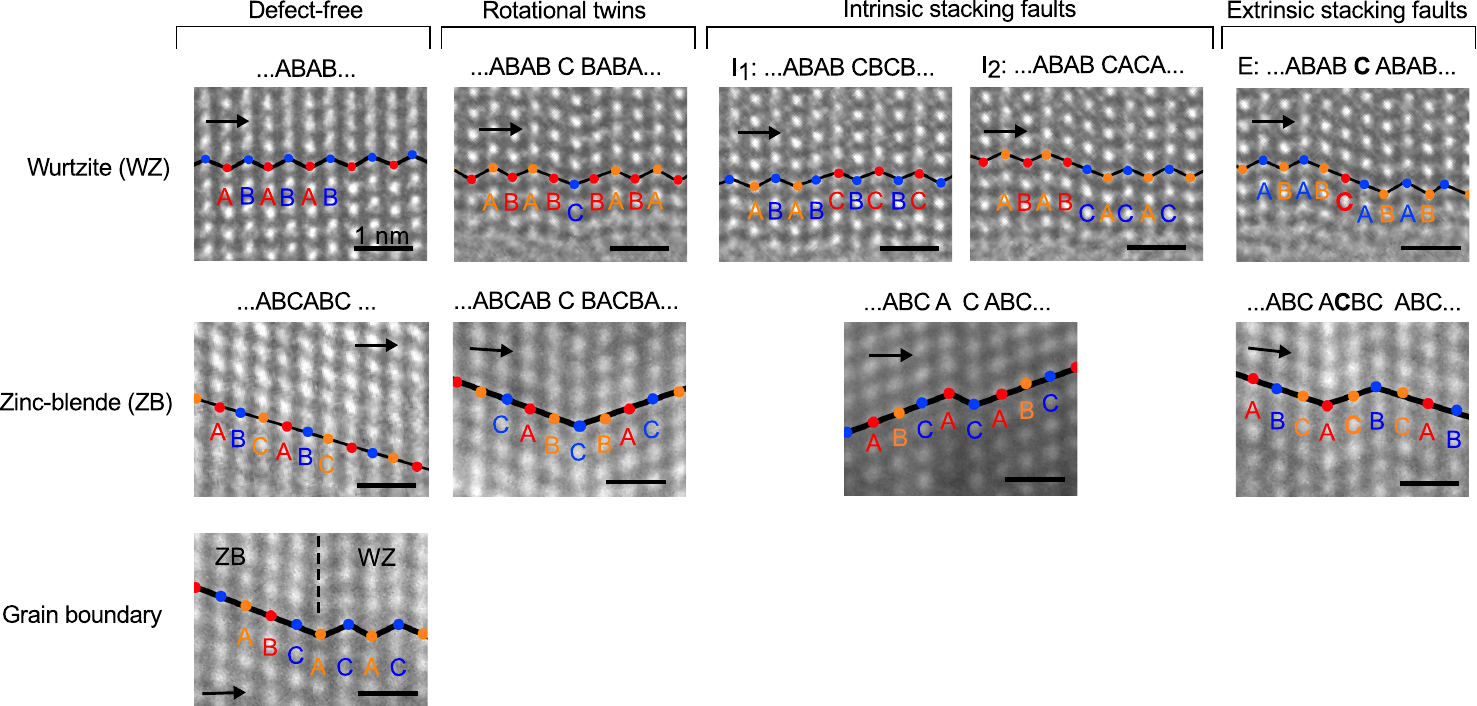}
\caption{TEM images of InAs and InAs$_{1-x}$Sb$_{x}$ nanowires showing defect-free regions of wurtzite and zinc-blende structures and the various planar defects that are observed in both structures, including rotational twins, stacking faults and grain boundaries. The black arrow indicates the growth direction. The scale bar is 1\,nm for all images.}
\label{fig:AllStructures}
\end{figure}

\clearpage
In order to quantify the distribution of polytypes and defects in the nanowires we have used high resolution TEM.
III-V semiconductor nanowires usually crystallize either in the hexagonal wurtzite (WZ) phase or in the cubic zinc-blende (ZB) phase. Other specific polytypes such as 4H and 6H have been reported on rare occasions in nanowires~\cite{Polytype4H, Polytype6H}. The zinc-blende structure is composed of interpenetrating face-centered cubic lattices corresponding to the stacking sequence ``...ABCABC...'' while the wurtzite structure consists of interpenetrating hexagonal lattices with a stacking ``...ABAB...''. Each letter in this stacking sequence corresponds to a bilayer (i.e. a pair of atomic layers) with vertically stacked group III and V atoms. 
The planar defects observed in the InAs and InAs$_{1-x}$Sb$_{x}$ nanowires can be classified in three main categories: rotational twins, stacking faults (SF) and grain boundaries. \ref{fig:AllStructures} details the stacking sequences of all encountered defects depending on the crystal structure. 
Rotational twins  occur when a segment of the structure is rotated by 60\textdegree around the <111> growth axis leading to a change in the stacking sequence which becomes a mirror image of the regular lattice.
The twin boundary occurs at the interface between the two regions with inverted stacking order. 
Stacking faults are partial displacements affecting the regular sequence in the stacking of the lattice planes. Intrinsic stacking faults result from a vacant plane while extrinsic stacking faults are due to the insertion of an extra plane in the sequence.
In the wurtzite structure, there are two types (I$_1$ and I$_2$) of intrinsic faults and one type (E) of extrinsic faults. The I$_2$ and E type faults can be treated as short sequences of cubic stacking with I$_2$ (i.e. ABCA) and E (i.e. ABCAB) corresponding to four and five cubic bilayers respectively. In the zinc-blende structure, there is one type of intrinsic stacking fault and one type of extrinsic fault . The cubic intrinsic stacking fault corresponds to 4 bilayers of hexagonal stacking. Finally, the grain boundary corresponds to a plane interconnecting two extended segments of single crystal phase, as seen in \ref{fig:AllStructures}.

\begin{figure}[ht!]
\includegraphics[width=\textwidth]{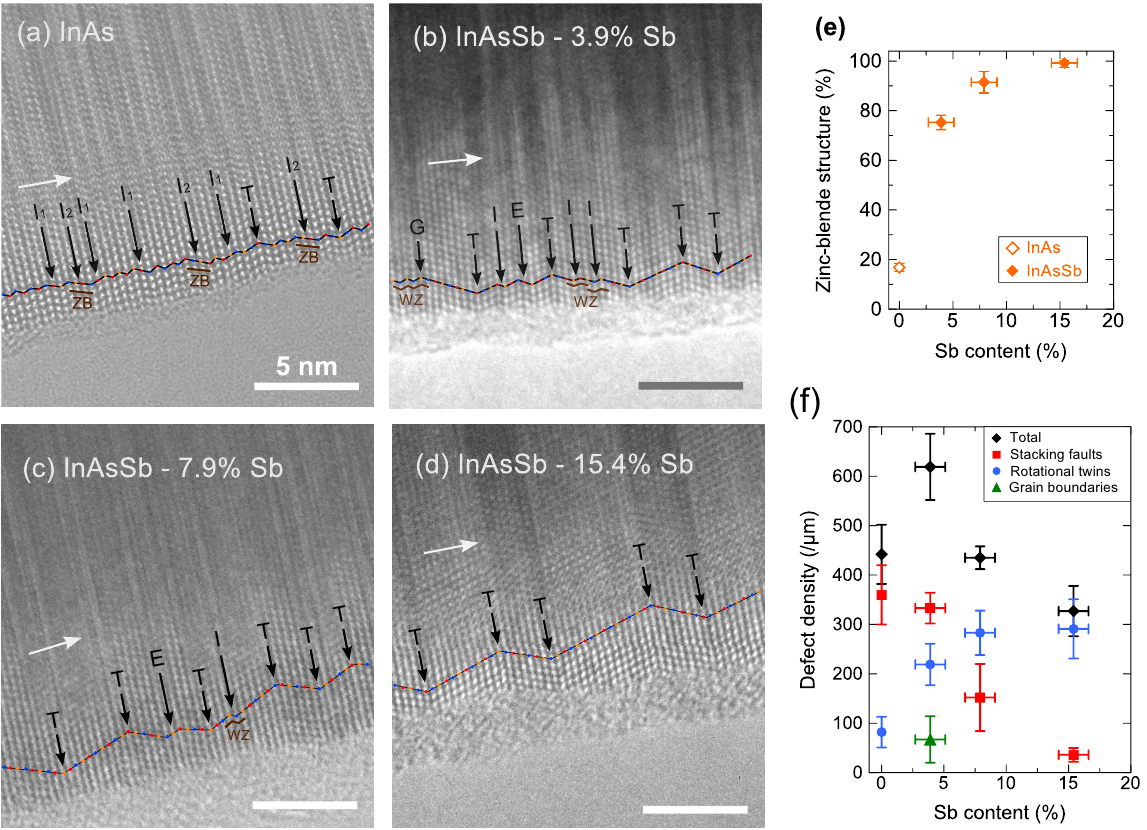}
\caption{(a--d) High resolution TEM characterization of catalyst-free InAs and InAs$_{1-x}$Sb$_{x}$ nanowires with increasing antimony content, all grown on silicon. Twins (T), stacking faults (intrinsic I and extrinsic E) and grain boundaries (G) are indicated in the figures. White arrows show the growth direction.
 The scale bar is 5\,nm for all images (higher resolution available in Supporting Information). (e) Percentage of zinc-blende structure in the InAs and InAs$_{1-x}$Sb$_{x}$ nanowires as a function of antimony content. (f) Defect density in the InAs and InAs$_{1-x}$Sb$_{x}$ nanowires as a function of antimony content. The black diamonds represent the total number of defects including twins, stacking faults and grain boundaries. Error bars represent plus or minus one standard deviation. }
\label{fig:MarinaX}
\end{figure}

\begin{figure}[ht!]
\includegraphics[]{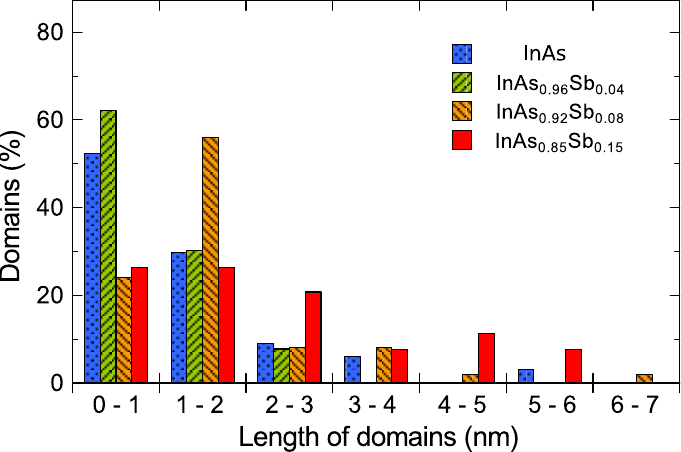}
\caption{Distribution of the length of the domains for catalyst-free InAs and InAs$_{1-x}$Sb$_{x}$ nanowires.}
\label{fig:LengthDistribution}
\end{figure}

Representative TEM images of nanowires with varying Sb content are shown in \ref{fig:MarinaX}(a-d). 
The crystal phase content of each group of nanowires was estimated from the observation of the bilayers in the TEM pictures. We follow the metrics proposed by Caroff et al.~\cite{MetricsCaroff} and consider that at least four consecutive bilayers following the same stacking sequence are required to assign a specific crystal phase to any segment. An illustration of a minimal ZB segment in an overall WZ structure could be ``...ABAB\textit{\underline{ABCA}}CACA...'' (i.e. an I$_2$ type fault).
Along with the crystal phase content, we estimated the planar defect density -- defects consisting of rotational twins, stacking faults and grain boundaries as detailed in \ref{fig:AllStructures} -- and the distribution of the domain length for each group of nanowires. We define a domain by a stacking of three or more bilayers without defects. \ref{fig:LengthDistribution} shows the distribution of the domain lengths for each group of nanowires.

InAs nanowires show a dominant wurtzite structure and only 17\% of the zinc-blende polytype on average. As seen in \ref{fig:MarinaX}(e), the incorporation of antimony into the InAs crystal completely changes the structure from wurtzite dominant to zinc-blende dominant, even for Sb content as low as 3.9\%. All InAs$_{1-x}$Sb$_{x}$ nanowires are predominantly composed of the zinc-blende polytype: from 75\% for InAs$_{0.96}$Sb$_{0.04}$ nanowires to 99\% for InAs$_{0.85}$Sb$_{0.15}$ nanowires.
InAs and InAs$_{0.96}$Sb$_{0.04}$ nanowires both exhibit a pronounced polytypism and very short domains: more than 82\% are less than 2\,nm. As commonly found in nanowires grown without the use of catalyst~\cite{MixedStructureInInAs01, UCF17, MixedStructureInInAs02}, the nanowires generally present a high density of planar defects -- here between 330 and 620 defects per micrometer. However, the distribution of defects into the different categories (twins, staking faults and grains) changes considerably with the antimony incorporation as seen in \ref{fig:MarinaX}(f). For increasing antimony content, the twin density increases from 80 twins per micrometer in the InAs nanowires to 300 twins per micrometer in the InAs$_{0.85}$Sb$_{0.15}$ nanowire. In parallel the stacking fault density drastically decreases, by up to one order of magnitude between the InAs nanowires (360 SF per micrometer) and the InAs$_{0.85}$Sb$_{0.15}$ nanowires (35 SF per micrometer). InAs$_{0.85}$Sb$_{0.15}$ nanowires also present much longer domains: only 33\% of the domains are less than 2\,nm.

These observations agree with theoretical models~\cite{0957-4484-23-9-095602} and previous works on gold-catalyzed InAs$_{1-x}$Sb$_{x}$ nanowires grown by MBE~\cite{InAsSbPerfectStructure02}, reporting variations in the structure and defect density as a function of antimony content.
The tendency of antimony compound semiconductors to form zinc-blende structure over wurtzite is generally attributed to the low ionicity of the atomic bonds~\cite{IonicityInSb, InSbNanowireArrays} and the growth kinetics~\cite{Dubrovskii_Glas2}. Although twin-free ZB phase has been achieved for antimony content in the 10\% range for gold-catalyzed InAs$_{1-x}$Sb$_{x}$ nanowires~\cite{InAsSbPerfectStructure02}, the self-catalyzed nature of our process might require an higher incorporation to reach the same crystal quality. Plissard et al.~\cite{GaAsSb_gold_free} reported a twin-free ZB phase in GaAs$_{0.66}$Sb$_{0.34}$ segments in GaAs/GaAsSb nanowires grown on silicon via a gold-free process.

\clearpage
\section{Electrical transport in InAs$_{1-x}$Sb$_{x}$ nanowires}
\vspace{0.3cm}

\begin{figure}[ht!]
\includegraphics[]{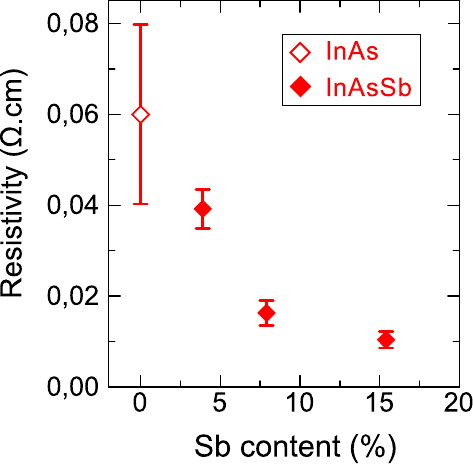}
\caption{Average nanowire resistivity at room temperature as a function of the antimony content, calculated assuming a bulk conductivity. Error bars represent plus or minus one standard deviation.}
\label{fig:ResistivityforDifferentGroups}
\end{figure}

Electrical characterizations were performed on the same three groups of InAs$_{1-x}$Sb$_{x}$ nanowires and on the pure InAs nanowires for reference.
Electrical transport can take place either in the bulk~\cite{yao:082103} or at the surface in an accumulation layer~\cite{WhereCarriersCome02}, and therefore various characteristics of the nanowires, such as resistivity or carrier concentration, scale with the diameter according to the type of transport~\cite{DiameterDependenceConductionInAsNWs}. 
Other diameter-dependent parameters include the gate capacitance (which is used to extract parameters from field-effect measurements) and the crystal structure. \footnote{Thelander et al.~\cite{PhaseMixingInAsNW} have shown that the dependence of crystal structure upon diameter is much less marked for MBE-grown than MOVPE-grown InAs nanowires.}
To isolate the influence of antimony incorporation on the transport properties, we restricted the study to nanowires with diameter between 110 and 120\,nm and assumed exclusively bulk conductivity in our calculations.
A two-dimensional model was also investigated (not presented here) and gave results very similar to the bulk conductivity model in terms of resistivity, carrier concentration and mobility. 
Similarly to our previous work~\cite{MY-PAPER-ContactResistance}, the contact resistance was found to be almost two orders of magnitude lower than the bulk nanowire resistance, for both InAs and InAs$_{1-x}$Sb$_{x}$ nanowires, and therefore negligible.
The resistivity of the different groups of nanowires was extracted using $\rho = \pi r^2 R/L$ where $r$ is the nanowire radius, $R$ is the nanowire resistance obtained from two-point measurements and $L$ is the distance between the contacts. As seen in~\ref{fig:ResistivityforDifferentGroups}, the resistivity of the nanowires significantly decreases upon incorporation of antimony. InAs$_{0.85}$Sb$_{0.15}$ nanowires show a resistivity six times smaller than InAs nanowires.

\vspace{1cm}

\begin{figure}[ht]
\includegraphics[]{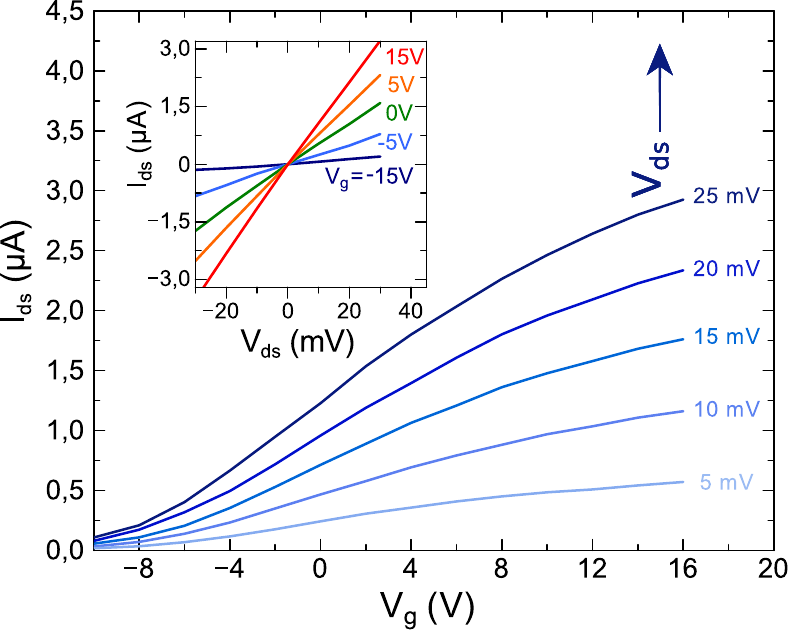}
\caption{Transfer characteristic of an InAs$_{0.92}$Sb$_{0.08}$ nanowire field-effect transistor at room temperature. Inset: Output characteristic of the same nanowire.}
\label{fig:NWFET-Transfer-Output}
\end{figure}

Field-effect measurements were performed to determine the nanowire mobility and carrier concentration. 
\ref{fig:NWFET-Transfer-Output}(Inset) shows typical $I_\text{ds}$ vs $V_\text{ds}$ output characteristic obtained for an InAs$_{0.92}$Sb$_{0.08}$ nanowire field-effect transistor at room temperature, with $I_\text{ds}$ and $V_\text{ds}$ respectively being the drain-source current and voltage, and $V_\text{g}$ the back-gate voltage applied through the silicon substrate. The increase of conductivity with the application of a positive gate voltage demonstrates the n-type behavior of the nanowires. All InAs and InAs$_{1-x}$Sb$_{x}$ nanowires were found to be n-type. \ref{fig:NWFET-Transfer-Output} shows the transfer  characteristic ($I_\text{ds}$ vs $V_\text{g}$ for different $V_\text{ds}$ ranging from 5 to 25\,mV) of the same device.
The field-effect mobility is estimated from $\mu = g_\text{m} (L^2/C)(1/V_\text{ds})$ 
where the transconductance $g_\text{m}$ equals $(dI_\text{ds})/(dV_\text{g})$ at a constant $V_\text{ds}$ and $L$ is the length of the channel, i.e. the distance between the contacts. $C$ is the gate-nanowire capacitance which is derived from the metallic cylinder on an infinite plane model:
$C = 2 \pi \epsilon_{0} \epsilon_\text{r} L  / \displaystyle \cosh ^{-1} [ (r+ t_\text{ox})/r ]  $~\cite{Smythe}.
Here $t_\text{ox}$ is the gate oxide thickness, $\epsilon_\text{r}$ is the relative dielectric constant of the dielectric material and $r$ is the nanowire radius. Details regarding the adjustments to this model to fit semiconductor nanowires can be found in the Supporting Information. The average estimated normalized capacitance was 5.3$\times 10^{-11}$\,F.m$^{-1}$ for the InAs and InAs$_{1-x}$Sb$_{x}$ devices. The peak field-effect mobility is obtained from the maximum value of $g_\text{m}$ with respect to $V_\text{g}$.
Applying a negative back-gate voltage reduces the number of free carriers in the n-type nanowire channel until it reaches a complete depletion at the pinch-off point. At this point, the induced charge $Q$ is equal to $CV_\text{th}$ where $V_\text{th}$ is the threshold voltage extracted from the transfer characteristic. $Q$ represents the charge in the bulk nanowire and is equal to $e \pi r^2 L n_\text{e}$ (once again assuming a bulk conductivity). 
The carrier concentration $n_\text{e}$ is therefore estimated by using $n_\text{e} = (C V_\text{th})/(e \pi r^2 L )$.

\vspace{0.7cm}

\begin{figure}[ht]
\includegraphics[]{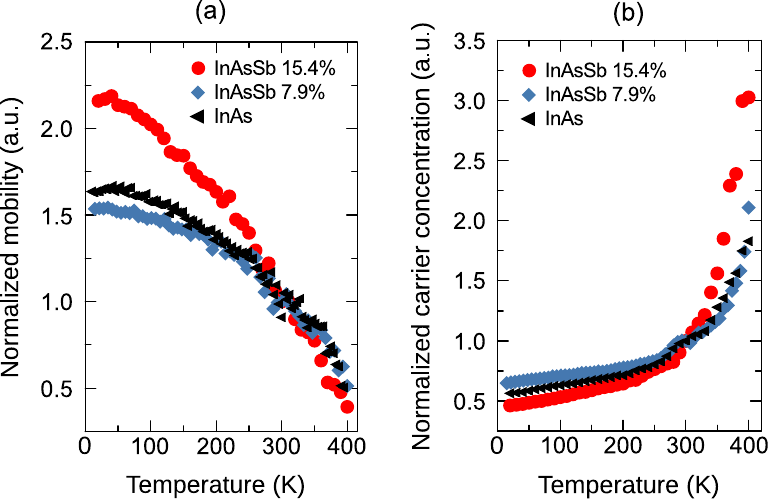}
\caption{Temperature dependence of (a) normalized peak field-effect mobility and (b) normalized carrier concentration of InAs and InAs$_{1-x}$Sb$_{x}$ nanowires with 7.9\% and 15.4\% antimony. The mobility and carrier concentration have been normalized with respect to the values obtained for each nanowire at room temperature.}
\label{fig:LowTempMeasurements}
\end{figure}

Field-effect measurements were performed between 400 and 10\,K on the InAs and InAs$_{1-x}$Sb$_{x}$ nanowires. 
The temperature-dependence of the extracted peak field-effect mobility and carrier concentration of individual nanowires is plotted in~\ref{fig:LowTempMeasurements}(a) and \ref{fig:LowTempMeasurements}(b) respectively. Following the classic semiconductor behavior, the carrier concentration increases with temperature as the extra thermal energy allows electrons to go to the conduction band. This effect is especially visible for InAs$_{1-x}$Sb$_{x}$ nanowires compared to InAs nanowires owing to their reduced band gap.
We observe an approximately linear enhancement in the mobility as the temperature goes down due to reduced phonon scattering.
The mobility starts to saturate at lower temperature at which point it becomes limited by scattering events related to the nanowire structure including planar defects~\cite{InAsNWOxideLayerThickness01}. 
Similar trends have been observed previously in InAs films~\cite{MobilityTemperatureBulkInAs}, InAs nanowires~\cite{InAsNWOxideLayerThickness01} and InAs-InP~\cite{MobilityTemperatureInNW02} and InAs-InAlAs~\cite{MobilityTemperatureInNW01} core-shell nanowires.

\begin{figure}[ht]
\includegraphics[]{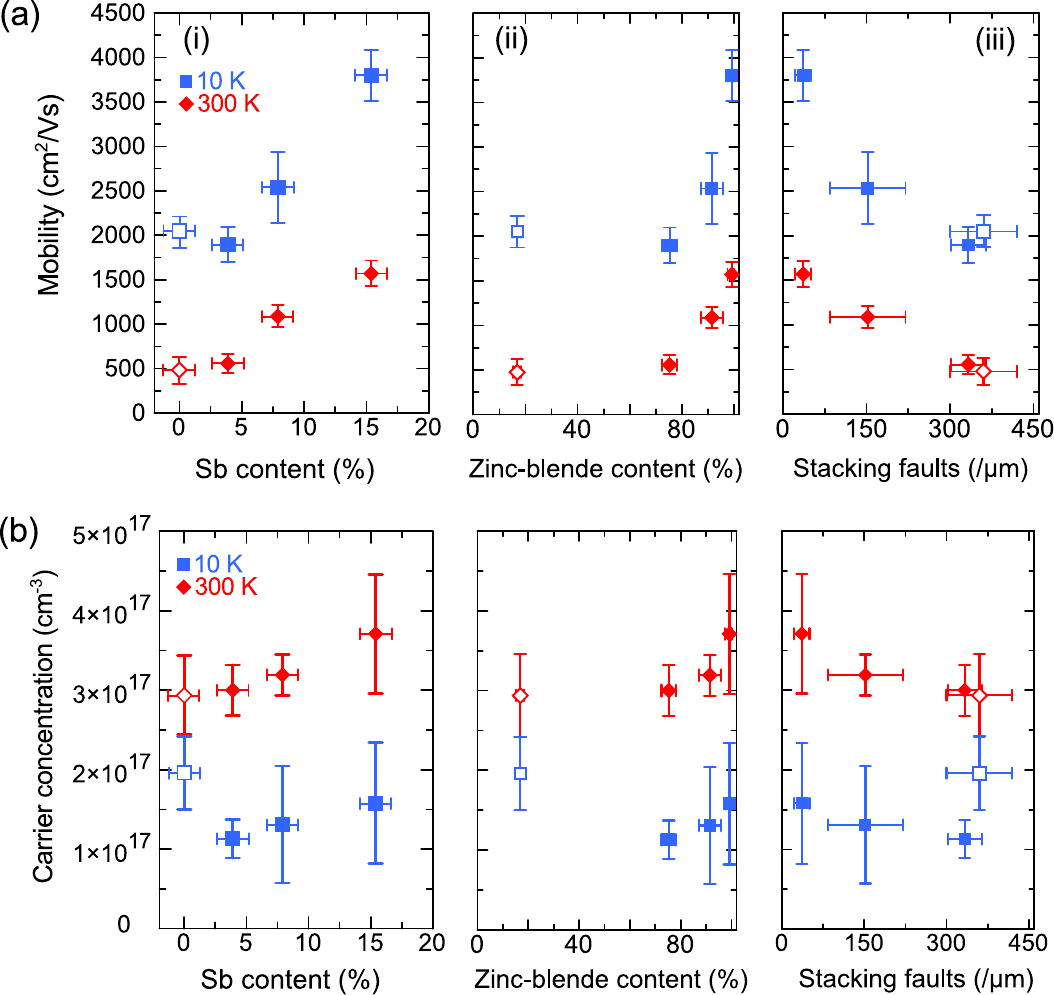}
\caption{(a) Nanowire mobility and (b) carrier concentration as a function of the (i) antimony content, (ii) zinc-blende content and (iii) stacking fault density. Red diamonds are room temperature data, blue squares are results obtained at 10\,K. Each point is an average of the results obtained for five to ten nanowires. InAs nanowires are indicated by empty symbols and InAs$_{1-x}$Sb$_{x}$ by filled symbols. Error bars represent plus or minus one standard deviation.}
\label{fig:Mobility-compar2}
\end{figure}

\clearpage
The average nanowire carrier concentration and peak mobility extracted from the field-effect measurements are displayed in \ref{fig:Mobility-compar2} as a function of antimony content, zinc-blende content and stacking fault density.
By taking into account the error bars, the carrier concentration shown in~\ref{fig:Mobility-compar2}(b) appears to be almost independent of the antimony incorporation. At room temperature, the average carrier concentration varies only very slightly with the Sb incorporation increasing from $2.9 \times 10^{17}$\,cm$^{-3}$ for InAs nanowires to $3.7 \times 10^{17}$\,cm$^{-3}$ for InAs$_{0.85}$Sb$_{0.15}$ nanowires.
This is probably attributable to the fact that the intrinsic carrier concentration is higher in bulk InSb~\cite{IntriniscCarrierConInSbReference} ($2\times10^{16}$\,cm$^{-3}$) than in InAs~\cite{IntriniscCarrierConInAs_Reference_Rogalski198935} ($8\times10^{14}$\,cm$^{-3}$).

The catalyst-free InAs nanowires exhibit an average mobility of $\sim$500\,cm$^2$/Vs at room temperature and $\sim$2000\,cm$^2$/Vs at low temperature. The room temperature mobility is lower than values usually reported for defect-free Au-assisted InAs nanowires grown by MBE or CBE (1000--2000\,cm$^2$/Vs range~\cite{0957-4484-21-20-205703, Vitiello, PhaseMixingInAsNW}) but is similar to the value reported for InAs nanowires with high defect density (500--750\,cm$^2$/Vs range~\cite{PhaseMixingInAsNW}).
Despite a major change in the type of structure from wurtzite-dominant to zinc-blende dominant, the InAs$_{0.96}$Sb$_{0.04}$ nanowires present an average mobility very similar to the InAs nanowires. Both groups of nanowires have a high number of stacking faults ($\sim$350\,/$\upmu$m) and especially small domain lengths. For higher antimony content, the mobility significantly increases. It reaches 1600\,cm$^2$/Vs at room temperature for InAs$_{0.85}$Sb$_{0.15}$ nanowires presenting longer domains, a low stacking fault density but relatively high twin density. At 10\,K, the mobility shows a similar trend, ranging from 1900\,cm$^2$/Vs for the InAs$_{0.96}$Sb$_{0.04}$ nanowires to 3750\,cm$^2$/Vs for InAs$_{0.85}$Sb$_{0.15}$ nanowires. 

Here we consider three different candidate mechanisms that could be causing this marked increase in mobility with antimony content: (i) the different intrinsic electrical properties of InAs$_{1-x}$Sb$_{x}$ as a function of Sb content; (ii) the different electrical properties of the WZ and ZB polytypes of InAs$_{1-x}$Sb$_{x}$; and (iii) the variation of the planar defect density. 
It seems unlikely that variation of the intrinsic properties with Sb content have a bearing on the mobility variation in our nanowires since the measured mobilities are two orders of magnitude lower than those of bulk InAs (40000\,cm$^2$/Vs) and InSb (77000\,cm$^2$/Vs)~\cite{CRCHandbook_Mobility}. In addition the fractional change in mobility as \textit{x} changes from zero to 0.15 for the nanowires is far greater than that for the corresponding bulk materials. We therefore eliminate the possibility that an intrinsic mechanism could account for the enhancement of mobility with Sb content. 
From \ref{fig:Mobility-compar2}(a,ii), the largest difference in the crystal structure polytype is observed between InAs and InAs$_{0.96}$Sb$_{0.04}$ nanowires but there is no significant difference in mobility between these two groups of nanowires. Therefore we can eliminate the possibility that different electrical properties of the WZ and ZB types account for the mobility change.
Finally we consider the defect density, meaning twins, grain boundaries and stacking faults. From the combined results presented in \ref{fig:MarinaX}(f) and \ref{fig:Mobility-compar2}(a,i), we observe that the mobility of the InAs$_{1-x}$Sb$_{x}$ nanowires increases with Sb content despite the increasing number of twin boundaries. This indicates that twins are not especially detrimental to the electron mobility. Grain boundaries are present in only one of the studied groups of nanowires thus we cannot determine how they might affect the mobility. Finally as seen in \ref{fig:Mobility-compar2}(a,iii), the field-effect mobility increases approximately linearly with the reduction of the stacking fault density. The change in mobility therefore seems to be dominated by the stacking fault density and the length of the domains. A stacking fault can lead to a change of structure with the introduction of short foreign domains (i.e. WZ domains in an overall ZB structure and vice versa), while the structure polytype does not change in the presence of a twin. This is in agreement with~\citeauthor{PhaseMixingInAsNW}~\cite{PhaseMixingInAsNW}, who observed in the context of heterocatalysed InAs nanowires that mixtures of extended wurtzite/zinc-blende segments affect the electrical properties more significantly than isolated twin planes.

These measurements strongly suggest that the mobility enhancement in the nanowires originates from the reduction and stacking fault density and the presence of longer domains, although other parameters not investigated in this work such as the surface roughness~\cite{0957-4484-24-37-375202} could also play a role.
Repeated type boundaries in the structure act as scattering sources for the carriers and create a strain in the structure modifying the carrier effective mass and therefore affecting the mobility~\cite{EnhancementMobilityStructure, EnhancementMobilityStructure02}.

\clearpage
\section{Conclusion}

In summary, we have grown InAs$_{1-x}$Sb$_{x}$ (0\,$\leq$\,x\,$\leq$\,0.15) nanowires on silicon substrates via a catalyst-free MBE process. Upon incorporation of antimony, we observed a drastic change in the crystal structure polytype from wurtzite-dominant in the InAs nanowires (17\% ZB) to almost pure zinc-blende in the InAs$_{0.85}$Sb$_{0.15}$ (99\% ZB). With increasing amounts of antimony, we observed a sharp decrease of the stacking fault density in the InAs$_{1-x}$Sb$_{x}$ nanowires and a significant increase of the domain length and field-effect mobility. The mobility was not significantly affected by the presence of defects in the form of twins.
 Through the adjustment of the growth conditions or antimony content, defect-free InAs$_{1-x}$Sb$_{x}$ nanowires with no polytypism may be achievable without the use of heterocatalytic nanoparticle seeds. This work is important for the development of functional nanowire devices integrated directly with silicon CMOS electronics.

\begin{acknowledgement}

We thank Kevin Lee for technical support.
This work was supported by EPSRC, grant reference EP/H005544/1.
The high resolution TEM measurements were carried out at University of St. Andrews by Prof. Wuzong Zhou and Miss Heather Greer. HRXRD measurements were performed at University College London.

\end{acknowledgement}


\begin{suppinfo}

\end{suppinfo}
\clearpage

\bibliography{mybiblio2}

\providecommand*{\mcitethebibliography}{\thebibliography}
\csname @ifundefined\endcsname{endmcitethebibliography}
{\let\endmcitethebibliography\endthebibliography}{}
\begin{mcitethebibliography}{53}
\providecommand*{\natexlab}[1]{#1}
\providecommand*{\mciteSetBstSublistMode}[1]{}
\providecommand*{\mciteSetBstMaxWidthForm}[2]{}
\providecommand*{\mciteBstWouldAddEndPuncttrue}
  {\def\EndOfBibitem{\unskip.}}
\providecommand*{\mciteBstWouldAddEndPunctfalse}
  {\let\EndOfBibitem\relax}
\providecommand*{\mciteSetBstMidEndSepPunct}[3]{}
\providecommand*{\mciteSetBstSublistLabelBeginEnd}[3]{}
\providecommand*{\EndOfBibitem}{}
\mciteSetBstSublistMode{f}
\mciteSetBstMaxWidthForm{subitem}{(\alph{mcitesubitemcount})}
\mciteSetBstSublistLabelBeginEnd{\mcitemaxwidthsubitemform\space}
{\relax}{\relax}

\bibitem[Xu et~al.(2013)Xu, Jiang, Qing, Mai, Zhang, and
  Lieber]{Lieber_NanoLett01}
Xu,~L.; Jiang,~Z.; Qing,~Q.; Mai,~L.; Zhang,~Q.; Lieber,~C.~M. \emph{Nano
  Lett.} \textbf{2013}, \emph{13}, 746--751\relax
\mciteBstWouldAddEndPuncttrue
\mciteSetBstMidEndSepPunct{\mcitedefaultmidpunct}
{\mcitedefaultendpunct}{\mcitedefaultseppunct}\relax
\EndOfBibitem
\bibitem[Wu et~al.(2012)Wu, Anttu, Xu, Samuelson, and
  Pistol]{Samuelson_NanoLett01}
Wu,~P.~M.; Anttu,~N.; Xu,~H.~Q.; Samuelson,~L.; Pistol,~M.-E. \emph{Nano Lett.}
  \textbf{2012}, \emph{12}, 1990--1995\relax
\mciteBstWouldAddEndPuncttrue
\mciteSetBstMidEndSepPunct{\mcitedefaultmidpunct}
{\mcitedefaultendpunct}{\mcitedefaultseppunct}\relax
\EndOfBibitem
\bibitem[Abay et~al.(2012)Abay, Nilsson, Wu, Xu, Wilson, and
  Delsing]{Abay_NanoLett01}
Abay,~S.; Nilsson,~H.; Wu,~F.; Xu,~H.; Wilson,~C.; Delsing,~P. \emph{Nano
  Lett.} \textbf{2012}, \emph{12}, 5622--5625\relax
\mciteBstWouldAddEndPuncttrue
\mciteSetBstMidEndSepPunct{\mcitedefaultmidpunct}
{\mcitedefaultendpunct}{\mcitedefaultseppunct}\relax
\EndOfBibitem
\bibitem[Huang et~al.(2002)Huang, Duan, Cui, and Lieber]{FET6}
Huang,~Y.; Duan,~X.; Cui,~Y.; Lieber,~C.~M. \emph{Nano Lett.} \textbf{2002},
  \emph{2}, 101--104\relax
\mciteBstWouldAddEndPuncttrue
\mciteSetBstMidEndSepPunct{\mcitedefaultmidpunct}
{\mcitedefaultendpunct}{\mcitedefaultseppunct}\relax
\EndOfBibitem
\bibitem[Qian et~al.(2004)Qian, Li, Gradecak, Wang, Barrelet, and Lieber]{LED3}
Qian,~F.; Li,~Y.; Gradecak,~S.; Wang,~D.; Barrelet,~C.~J.; Lieber,~C.~M.
  \emph{Nano Lett.} \textbf{2004}, \emph{4}, 1975--1979\relax
\mciteBstWouldAddEndPuncttrue
\mciteSetBstMidEndSepPunct{\mcitedefaultmidpunct}
{\mcitedefaultendpunct}{\mcitedefaultseppunct}\relax
\EndOfBibitem
\bibitem[Isakov et~al.(2013)Isakov, Panfilova, Sourribes, Tileli, Porter, and
  Warburton]{IVAN-PAPER}
Isakov,~I.; Panfilova,~M.; Sourribes,~M. J.~L.; Tileli,~V.; Porter,~A.~E.;
  Warburton,~P.~A. \emph{Nanotechnology} \textbf{2013}, \emph{24}, 085707\relax
\mciteBstWouldAddEndPuncttrue
\mciteSetBstMidEndSepPunct{\mcitedefaultmidpunct}
{\mcitedefaultendpunct}{\mcitedefaultseppunct}\relax
\EndOfBibitem
\bibitem[Tomioka et~al.(2010)Tomioka, Motohisa, Hara, Hiruma, and
  Fukui]{AlGaAs_01}
Tomioka,~K.; Motohisa,~J.; Hara,~S.; Hiruma,~K.; Fukui,~T. \emph{Nano Lett.}
  \textbf{2010}, \emph{10}, 1639--1644\relax
\mciteBstWouldAddEndPuncttrue
\mciteSetBstMidEndSepPunct{\mcitedefaultmidpunct}
{\mcitedefaultendpunct}{\mcitedefaultseppunct}\relax
\EndOfBibitem
\bibitem[Chuang et~al.(2011)Chuang, Moewe, Ng, Tran, Crankshaw, Chen, Ko, and
  Chang-Hasnain]{LatticeMismatchNW02}
Chuang,~L.~C.; Moewe,~M.; Ng,~K.~W.; Tran,~T.-T.~D.; Crankshaw,~S.; Chen,~R.;
  Ko,~W.~S.; Chang-Hasnain,~C. \emph{Appl. Phys. Lett.} \textbf{2011},
  \emph{98}, 123101\relax
\mciteBstWouldAddEndPuncttrue
\mciteSetBstMidEndSepPunct{\mcitedefaultmidpunct}
{\mcitedefaultendpunct}{\mcitedefaultseppunct}\relax
\EndOfBibitem
\bibitem[Glas(2006)]{Glas1}
Glas,~F. \emph{Phys. Rev. B} \textbf{2006}, \emph{74}, 121302\relax
\mciteBstWouldAddEndPuncttrue
\mciteSetBstMidEndSepPunct{\mcitedefaultmidpunct}
{\mcitedefaultendpunct}{\mcitedefaultseppunct}\relax
\EndOfBibitem
\bibitem[Tomioka et~al.(2011)Tomioka, Tanaka, Hara, Hiruma, and
  Fukui]{SelectiveArea8}
Tomioka,~K.; Tanaka,~T.; Hara,~S.; Hiruma,~K.; Fukui,~T. \emph{IEEE Journal on
  Selected Topics in Quantum Electronics} \textbf{2011}, \emph{17},
  1112--1129\relax
\mciteBstWouldAddEndPuncttrue
\mciteSetBstMidEndSepPunct{\mcitedefaultmidpunct}
{\mcitedefaultendpunct}{\mcitedefaultseppunct}\relax
\EndOfBibitem
\bibitem[Tomioka et~al.(2008)Tomioka, Motohisa, Hara, and
  Fukui]{MixedStructureInInAs01}
Tomioka,~K.; Motohisa,~J.; Hara,~S.; Fukui,~T. \emph{Nano Lett.} \textbf{2008},
  \emph{8}, 3475--3480\relax
\mciteBstWouldAddEndPuncttrue
\mciteSetBstMidEndSepPunct{\mcitedefaultmidpunct}
{\mcitedefaultendpunct}{\mcitedefaultseppunct}\relax
\EndOfBibitem
\bibitem[Brotherton and Lowther(1980)]{PhysRevLett.44.606}
Brotherton,~S.~D.; Lowther,~J.~E. \emph{Phys. Rev. Lett.} \textbf{1980},
  \emph{44}, 606--609\relax
\mciteBstWouldAddEndPuncttrue
\mciteSetBstMidEndSepPunct{\mcitedefaultmidpunct}
{\mcitedefaultendpunct}{\mcitedefaultseppunct}\relax
\EndOfBibitem
\bibitem[Allen et~al.(2008)Allen, Hemesath, Perea, Lensch-Falk, Li, Yin, Gass,
  Wang, Bleloch, Palmer, and Lauhon]{GoldInSilicon02}
Allen,~J.~E.; Hemesath,~E.~R.; Perea,~D.~E.; Lensch-Falk,~J.~L.; Li,~Z.~Y.;
  Yin,~F.; Gass,~M.~H.; Wang,~P.; Bleloch,~A.~L.; Palmer,~R.~E.; Lauhon,~L.~J.
  \emph{Nature Nanotechnology} \textbf{2008}, \emph{3}, 168--173\relax
\mciteBstWouldAddEndPuncttrue
\mciteSetBstMidEndSepPunct{\mcitedefaultmidpunct}
{\mcitedefaultendpunct}{\mcitedefaultseppunct}\relax
\EndOfBibitem
\bibitem[Bar-Sadan et~al.(2012)Bar-Sadan, Barthel, Shtrikman, and
  Houben]{GoldInSilicon03}
Bar-Sadan,~M.; Barthel,~J.; Shtrikman,~H.; Houben,~L. \emph{Nano Lett.}
  \textbf{2012}, \emph{12}, 2352--2356\relax
\mciteBstWouldAddEndPuncttrue
\mciteSetBstMidEndSepPunct{\mcitedefaultmidpunct}
{\mcitedefaultendpunct}{\mcitedefaultseppunct}\relax
\EndOfBibitem
\bibitem[Ford et~al.(2009)Ford, Ho, Chueh, Tseng, Fan, Guo, Bokor, and
  Javey]{InAsNWOxideLayerThickness01}
Ford,~A.~C.; Ho,~J.~C.; Chueh,~Y.-L.; Tseng,~Y.-C.; Fan,~Z.; Guo,~J.;
  Bokor,~J.; Javey,~A. \emph{Nano Lett.} \textbf{2009}, \emph{9},
  360--365\relax
\mciteBstWouldAddEndPuncttrue
\mciteSetBstMidEndSepPunct{\mcitedefaultmidpunct}
{\mcitedefaultendpunct}{\mcitedefaultseppunct}\relax
\EndOfBibitem
\bibitem[Mandl et~al.(2010)Mandl, Stangl, Hilner, Zakharov, Hillerich, Dey,
  Samuelson, Bauer, Deppert, and Mikkelsen]{MandlGrowth}
Mandl,~B.; Stangl,~J.; Hilner,~E.; Zakharov,~A.~A.; Hillerich,~K.; Dey,~A.~W.;
  Samuelson,~L.; Bauer,~G.; Deppert,~K.; Mikkelsen,~A. \emph{Nano Lett.}
  \textbf{2010}, \emph{10}, 4443--4449\relax
\mciteBstWouldAddEndPuncttrue
\mciteSetBstMidEndSepPunct{\mcitedefaultmidpunct}
{\mcitedefaultendpunct}{\mcitedefaultseppunct}\relax
\EndOfBibitem
\bibitem[Mandl et~al.(2006)Mandl, Stangl, M\aa{}rtensson, Mikkelsen, Eriksson,
  Karlsson, Bauer, Samuelson, and Seifert]{MixedStructureInInAs02}
Mandl,~B.; Stangl,~J.; M\aa{}rtensson,~T.; Mikkelsen,~A.; Eriksson,~J.;
  Karlsson,~L.~S.; Bauer,~G.; Samuelson,~L.; Seifert,~W. \emph{Nano Lett.}
  \textbf{2006}, \emph{6}, 1817--1821\relax
\mciteBstWouldAddEndPuncttrue
\mciteSetBstMidEndSepPunct{\mcitedefaultmidpunct}
{\mcitedefaultendpunct}{\mcitedefaultseppunct}\relax
\EndOfBibitem
\bibitem[Koblm\"uller et~al.(2010)Koblm\"uller, Hertenberger, Vizbaras,
  Bichler, Bao, Zhang, and Abstreiter]{SelfGrowth2}
Koblm\"uller,~G.; Hertenberger,~S.; Vizbaras,~K.; Bichler,~M.; Bao,~F.;
  Zhang,~J.~.; Abstreiter,~G. \emph{Nanotechnology} \textbf{2010}, \emph{21},
  year\relax
\mciteBstWouldAddEndPuncttrue
\mciteSetBstMidEndSepPunct{\mcitedefaultmidpunct}
{\mcitedefaultendpunct}{\mcitedefaultseppunct}\relax
\EndOfBibitem
\bibitem[Gao et~al.(2009)Gao, Woo, Liang, Pozuelo, Prikhodko, Jackson, Goel,
  Hudait, Huffaker, Goorsky, Kodambaka, and Hicks]{SelfGrowth7}
Gao,~L.; Woo,~R.~L.; Liang,~B.; Pozuelo,~M.; Prikhodko,~S.; Jackson,~M.;
  Goel,~N.; Hudait,~M.~K.; Huffaker,~D.~L.; Goorsky,~M.~S.; Kodambaka,~S.;
  Hicks,~R.~F. \emph{Nano Lett.} \textbf{2009}, \emph{9}, 2223--2228\relax
\mciteBstWouldAddEndPuncttrue
\mciteSetBstMidEndSepPunct{\mcitedefaultmidpunct}
{\mcitedefaultendpunct}{\mcitedefaultseppunct}\relax
\EndOfBibitem
\bibitem[Zhang et~al.(2013)Zhang, Aagesen, Holm, J\o{}rgensen, Wu, and
  Liu]{SelfGrowth6-Huyiun}
Zhang,~Y.; Aagesen,~M.; Holm,~J.~V.; J\o{}rgensen,~H.~I.; Wu,~J.; Liu,~H.
  \emph{Nano Lett.} \textbf{2013}, \emph{13}, 3897--3902\relax
\mciteBstWouldAddEndPuncttrue
\mciteSetBstMidEndSepPunct{\mcitedefaultmidpunct}
{\mcitedefaultendpunct}{\mcitedefaultseppunct}\relax
\EndOfBibitem
\bibitem[Wei et~al.(2009)Wei, Bao, Soci, Ding, Wang, and
  Wang]{InAsWithCatalyst_01}
Wei,~W.; Bao,~X.-Y.; Soci,~C.; Ding,~Y.; Wang,~Z.-L.; Wang,~D. \emph{Nano
  Lett.} \textbf{2009}, \emph{9}, 2926--2934\relax
\mciteBstWouldAddEndPuncttrue
\mciteSetBstMidEndSepPunct{\mcitedefaultmidpunct}
{\mcitedefaultendpunct}{\mcitedefaultseppunct}\relax
\EndOfBibitem
\bibitem[Thelander et~al.(2011)Thelander, Caroff, Plissard, Dey, and
  Dick]{PhaseMixingInAsNW}
Thelander,~C.; Caroff,~P.; Plissard,~S.; Dey,~A.~W.; Dick,~K.~A. \emph{Nano
  Lett.} \textbf{2011}, \emph{11}, 2424--2429\relax
\mciteBstWouldAddEndPuncttrue
\mciteSetBstMidEndSepPunct{\mcitedefaultmidpunct}
{\mcitedefaultendpunct}{\mcitedefaultseppunct}\relax
\EndOfBibitem
\bibitem[Rogalski(2003)]{ReviewInfraredDetector}
Rogalski,~A. \emph{Progress in Quantum Electronics} \textbf{2003}, \emph{27},
  59 -- 210\relax
\mciteBstWouldAddEndPuncttrue
\mciteSetBstMidEndSepPunct{\mcitedefaultmidpunct}
{\mcitedefaultendpunct}{\mcitedefaultseppunct}\relax
\EndOfBibitem
\bibitem[Thelander et~al.(2012)Thelander, Caroff, Plissard, and
  Dick]{thelander:232105}
Thelander,~C.; Caroff,~P.; Plissard,~S.; Dick,~K.~A. \emph{Appl. Phys. Lett.}
  \textbf{2012}, \emph{100}, 232105\relax
\mciteBstWouldAddEndPuncttrue
\mciteSetBstMidEndSepPunct{\mcitedefaultmidpunct}
{\mcitedefaultendpunct}{\mcitedefaultseppunct}\relax
\EndOfBibitem
\bibitem[Sourribes et~al.(2013)Sourribes, Isakov, Panfilova, and
  Warburton]{MY-PAPER-ContactResistance}
Sourribes,~M. J.~L.; Isakov,~I.; Panfilova,~M.; Warburton,~P.~A.
  \emph{Nanotechnology} \textbf{2013}, \emph{24}, 045703\relax
\mciteBstWouldAddEndPuncttrue
\mciteSetBstMidEndSepPunct{\mcitedefaultmidpunct}
{\mcitedefaultendpunct}{\mcitedefaultseppunct}\relax
\EndOfBibitem
\bibitem[Vegard(1921)]{VegardsLaw}
Vegard,~L. \emph{Zeitschrift f\"ur Physik} \textbf{1921}, \emph{5},
  17--26\relax
\mciteBstWouldAddEndPuncttrue
\mciteSetBstMidEndSepPunct{\mcitedefaultmidpunct}
{\mcitedefaultendpunct}{\mcitedefaultseppunct}\relax
\EndOfBibitem
\bibitem[Tomioka et~al.(2007)Tomioka, Mohan, Noborisaka, Hara, Motohisa, and
  Fukui]{SelectiveArea2}
Tomioka,~K.; Mohan,~P.; Noborisaka,~J.; Hara,~S.; Motohisa,~J.; Fukui,~T.
  \emph{Journal of Crystal Growth} \textbf{2007}, \emph{298}, 644--647\relax
\mciteBstWouldAddEndPuncttrue
\mciteSetBstMidEndSepPunct{\mcitedefaultmidpunct}
{\mcitedefaultendpunct}{\mcitedefaultseppunct}\relax
\EndOfBibitem
\bibitem[Dheeraj et~al.(2008)Dheeraj, Patriarche, Zhou, Hoang, Moses,
  Gr\o{}nsberg, van Helvoort, Fimland, and Weman]{Polytype4H}
Dheeraj,~D.~L.; Patriarche,~G.; Zhou,~H.; Hoang,~T.~B.; Moses,~A.~F.;
  Gr\o{}nsberg,~S.; van Helvoort,~A. T.~J.; Fimland,~B.-O.; Weman,~H.
  \emph{Nano Lett.} \textbf{2008}, \emph{8}, 4459--4463\relax
\mciteBstWouldAddEndPuncttrue
\mciteSetBstMidEndSepPunct{\mcitedefaultmidpunct}
{\mcitedefaultendpunct}{\mcitedefaultseppunct}\relax
\EndOfBibitem
\bibitem[Dubrovskii and Sibirev(2008)]{Polytype6H}
Dubrovskii,~V.~G.; Sibirev,~N.~V. \emph{Phys. Rev. B} \textbf{2008}, \emph{77},
  035414\relax
\mciteBstWouldAddEndPuncttrue
\mciteSetBstMidEndSepPunct{\mcitedefaultmidpunct}
{\mcitedefaultendpunct}{\mcitedefaultseppunct}\relax
\EndOfBibitem
\bibitem[Caroff et~al.(2011)Caroff, Bolinsson, and Johansson]{MetricsCaroff}
Caroff,~P.; Bolinsson,~J.; Johansson,~J. \emph{Selected Topics in Quantum
  Electronics, IEEE Journal of} \textbf{2011}, \emph{17}, 829--846\relax
\mciteBstWouldAddEndPuncttrue
\mciteSetBstMidEndSepPunct{\mcitedefaultmidpunct}
{\mcitedefaultendpunct}{\mcitedefaultseppunct}\relax
\EndOfBibitem
\bibitem[Frielinghaus et~al.(2012)Frielinghaus, Fl\"{o}hr, Sladek, Weirich,
  Trellenkamp, Hardtdegen, Sch\"{a}pers, Schneider, and Meyer]{UCF17}
Frielinghaus,~R.; Fl\"{o}hr,~K.; Sladek,~K.; Weirich,~T.~E.; Trellenkamp,~S.;
  Hardtdegen,~H.; Sch\"{a}pers,~T.; Schneider,~C.~M.; Meyer,~C. \emph{Appl.
  Phys. Lett.} \textbf{2012}, \emph{101}, 062104\relax
\mciteBstWouldAddEndPuncttrue
\mciteSetBstMidEndSepPunct{\mcitedefaultmidpunct}
{\mcitedefaultendpunct}{\mcitedefaultseppunct}\relax
\EndOfBibitem
\bibitem[Lugani et~al.(2012)Lugani, Ercolani, Sorba, Sibirev, Timofeeva, and
  Dubrovskii]{0957-4484-23-9-095602}
Lugani,~L.; Ercolani,~D.; Sorba,~L.; Sibirev,~N.~V.; Timofeeva,~M.~A.;
  Dubrovskii,~V.~G. \emph{Nanotechnology} \textbf{2012}, \emph{23},
  095602\relax
\mciteBstWouldAddEndPuncttrue
\mciteSetBstMidEndSepPunct{\mcitedefaultmidpunct}
{\mcitedefaultendpunct}{\mcitedefaultseppunct}\relax
\EndOfBibitem
\bibitem[Xu et~al.(2012)Xu, Dick, Plissard, Nguyen, Makoudi, Berthe, Nys,
  Wallart, Grandidier, and Caroff]{InAsSbPerfectStructure02}
Xu,~T.; Dick,~K.~A.; Plissard,~S.; Nguyen,~T.~H.; Makoudi,~Y.; Berthe,~M.;
  Nys,~J.-P.; Wallart,~X.; Grandidier,~B.; Caroff,~P. \emph{Nanotechnology}
  \textbf{2012}, \emph{23}, 095702\relax
\mciteBstWouldAddEndPuncttrue
\mciteSetBstMidEndSepPunct{\mcitedefaultmidpunct}
{\mcitedefaultendpunct}{\mcitedefaultseppunct}\relax
\EndOfBibitem
\bibitem[Christensen et~al.(1987)Christensen, Satpathy, and
  Pawlowska]{IonicityInSb}
Christensen,~N.~E.; Satpathy,~S.; Pawlowska,~Z. \emph{Phys. Rev. B}
  \textbf{1987}, \emph{36}, 1032--1050\relax
\mciteBstWouldAddEndPuncttrue
\mciteSetBstMidEndSepPunct{\mcitedefaultmidpunct}
{\mcitedefaultendpunct}{\mcitedefaultseppunct}\relax
\EndOfBibitem
\bibitem[Vogel et~al.(2011)Vogel, de~Boor, Wittemann, Mensah, Werner, and
  Schmidt]{InSbNanowireArrays}
Vogel,~A.~T.; de~Boor,~J.; Wittemann,~J.~V.; Mensah,~S.~L.; Werner,~P.;
  Schmidt,~V. \emph{Cryst. Growth Des.} \textbf{2011}, \emph{11},
  1896--1900\relax
\mciteBstWouldAddEndPuncttrue
\mciteSetBstMidEndSepPunct{\mcitedefaultmidpunct}
{\mcitedefaultendpunct}{\mcitedefaultseppunct}\relax
\EndOfBibitem
\bibitem[Dubrovskii et~al.(2008)Dubrovskii, Sibirev, Harmand, and
  Glas]{Dubrovskii_Glas2}
Dubrovskii,~V.~G.; Sibirev,~N.~V.; Harmand,~J.~C.; Glas,~F. \emph{Phys. Rev. B}
  \textbf{2008}, \emph{78}, 235301\relax
\mciteBstWouldAddEndPuncttrue
\mciteSetBstMidEndSepPunct{\mcitedefaultmidpunct}
{\mcitedefaultendpunct}{\mcitedefaultseppunct}\relax
\EndOfBibitem
\bibitem[Plissard et~al.(2010)Plissard, Dick, Wallart, and
  Caroff]{GaAsSb_gold_free}
Plissard,~S.; Dick,~K.~A.; Wallart,~X.; Caroff,~P. \emph{Applied Physics
  Letters} \textbf{2010}, \emph{96}, 121901\relax
\mciteBstWouldAddEndPuncttrue
\mciteSetBstMidEndSepPunct{\mcitedefaultmidpunct}
{\mcitedefaultendpunct}{\mcitedefaultseppunct}\relax
\EndOfBibitem
\bibitem[Yao et~al.(2012)Yao, G\"unel, Bl\"omers, Weis, Chi, Lu, Liu,
  Gr\"utzmacher, and Sch\"apers]{yao:082103}
Yao,~H.; G\"unel,~H.~Y.; Bl\"omers,~C.; Weis,~K.; Chi,~J.; Lu,~J.~G.; Liu,~J.;
  Gr\"utzmacher,~D.; Sch\"apers,~T. \emph{Appl. Phys. Lett.} \textbf{2012},
  \emph{101}, 082103\relax
\mciteBstWouldAddEndPuncttrue
\mciteSetBstMidEndSepPunct{\mcitedefaultmidpunct}
{\mcitedefaultendpunct}{\mcitedefaultseppunct}\relax
\EndOfBibitem
\bibitem[Werner et~al.(2009)Werner, Limbach, Carsten, Denker, Malindretos, and
  Rizzi]{WhereCarriersCome02}
Werner,~F.; Limbach,~F.; Carsten,~M.; Denker,~C.; Malindretos,~J.; Rizzi,~A.
  \emph{Nano Lett.} \textbf{2009}, \emph{9}, 1567--1571, PMID: 19290610\relax
\mciteBstWouldAddEndPuncttrue
\mciteSetBstMidEndSepPunct{\mcitedefaultmidpunct}
{\mcitedefaultendpunct}{\mcitedefaultseppunct}\relax
\EndOfBibitem
\bibitem[Scheffler et~al.(2009)Scheffler, Nadj-Perge, Kouwenhoven, Borgstr\"om,
  and Bakkers]{DiameterDependenceConductionInAsNWs}
Scheffler,~M.; Nadj-Perge,~S.; Kouwenhoven,~L.~P.; Borgstr\"om,~M.~T.;
  Bakkers,~E. P. A.~M. \emph{J. Appl. Phys.} \textbf{2009}, \emph{106},
  124303\relax
\mciteBstWouldAddEndPuncttrue
\mciteSetBstMidEndSepPunct{\mcitedefaultmidpunct}
{\mcitedefaultendpunct}{\mcitedefaultseppunct}\relax
\EndOfBibitem
\bibitem[Smythe(1989)]{Smythe}
Smythe,~W.~R. \emph{Static and Dynamic Electricity};
\newblock Hemisphere Publishing Corporation, 1989\relax
\mciteBstWouldAddEndPuncttrue
\mciteSetBstMidEndSepPunct{\mcitedefaultmidpunct}
{\mcitedefaultendpunct}{\mcitedefaultseppunct}\relax
\EndOfBibitem
\bibitem[Rode(1971)]{MobilityTemperatureBulkInAs}
Rode,~D.~L. \emph{Phys. Rev. B} \textbf{1971}, \emph{3}, 3287--3299\relax
\mciteBstWouldAddEndPuncttrue
\mciteSetBstMidEndSepPunct{\mcitedefaultmidpunct}
{\mcitedefaultendpunct}{\mcitedefaultseppunct}\relax
\EndOfBibitem
\bibitem[Jiang et~al.(2007)Jiang, Xiong, Nam, Qian, Li, and
  Lieber]{MobilityTemperatureInNW02}
Jiang,~X.; Xiong,~Q.; Nam,~S.; Qian,~F.; Li,~Y.; Lieber,~C.~M. \emph{Nano
  Lett.} \textbf{2007}, \emph{7}, 3214--3218\relax
\mciteBstWouldAddEndPuncttrue
\mciteSetBstMidEndSepPunct{\mcitedefaultmidpunct}
{\mcitedefaultendpunct}{\mcitedefaultseppunct}\relax
\EndOfBibitem
\bibitem[Holloway et~al.(2013)Holloway, Song, Haapamaki, LaPierre, and
  Baugh]{MobilityTemperatureInNW01}
Holloway,~G.~W.; Song,~Y.; Haapamaki,~C.~M.; LaPierre,~R.~R.; Baugh,~J.
  \emph{Appl. Phys. Lett.} \textbf{2013}, \emph{102}, 043115\relax
\mciteBstWouldAddEndPuncttrue
\mciteSetBstMidEndSepPunct{\mcitedefaultmidpunct}
{\mcitedefaultendpunct}{\mcitedefaultseppunct}\relax
\EndOfBibitem
\bibitem[Oszwa\l{}ldowski and Zimpel(1988)]{IntriniscCarrierConInSbReference}
Oszwa\l{}ldowski,~M.; Zimpel,~M. \emph{Journal of Physics and Chemistry of
  Solids} \textbf{1988}, \emph{49}, 1179--1185\relax
\mciteBstWouldAddEndPuncttrue
\mciteSetBstMidEndSepPunct{\mcitedefaultmidpunct}
{\mcitedefaultendpunct}{\mcitedefaultseppunct}\relax
\EndOfBibitem
\bibitem[Rogalski and
  J\'o{}\'z{}wikowski(1989)]{IntriniscCarrierConInAs_Reference_Rogalski198935}
Rogalski,~A.; J\'o{}\'z{}wikowski,~K. \emph{Infrared Physics} \textbf{1989},
  \emph{29}, 35 -- 42\relax
\mciteBstWouldAddEndPuncttrue
\mciteSetBstMidEndSepPunct{\mcitedefaultmidpunct}
{\mcitedefaultendpunct}{\mcitedefaultseppunct}\relax
\EndOfBibitem
\bibitem[Thelander et~al.(2010)Thelander, Dick, Borgstr\"o{}m, Fr\"o{}berg,
  Caroff, Nilsson, and Samuelson]{0957-4484-21-20-205703}
Thelander,~C.; Dick,~K.~A.; Borgstr\"o{}m,~M.~T.; Fr\"o{}berg,~L.~E.;
  Caroff,~P.; Nilsson,~H.~A.; Samuelson,~L. \emph{Nanotechnology}
  \textbf{2010}, \emph{21}, 205703\relax
\mciteBstWouldAddEndPuncttrue
\mciteSetBstMidEndSepPunct{\mcitedefaultmidpunct}
{\mcitedefaultendpunct}{\mcitedefaultseppunct}\relax
\EndOfBibitem
\bibitem[Vitiello et~al.(2012)Vitiello, Coquillat, Viti, Ercolani, Teppe,
  Pitanti, Beltram, Sorba, Knap, and Tredicucci]{Vitiello}
Vitiello,~M.~S.; Coquillat,~D.; Viti,~L.; Ercolani,~D.; Teppe,~F.; Pitanti,~A.;
  Beltram,~F.; Sorba,~L.; Knap,~W.; Tredicucci,~A. \emph{Nano Letters}
  \textbf{2012}, \emph{12}, 96--101\relax
\mciteBstWouldAddEndPuncttrue
\mciteSetBstMidEndSepPunct{\mcitedefaultmidpunct}
{\mcitedefaultendpunct}{\mcitedefaultseppunct}\relax
\EndOfBibitem
\bibitem[Lide(1985)]{CRCHandbook_Mobility}
Lide,~D.~R. \emph{Handbook of Chemistry and Physics};
\newblock CRC Press Inc; 66th edition, 1985;
\newblock pp E--99\relax
\mciteBstWouldAddEndPuncttrue
\mciteSetBstMidEndSepPunct{\mcitedefaultmidpunct}
{\mcitedefaultendpunct}{\mcitedefaultseppunct}\relax
\EndOfBibitem
\bibitem[Wang et~al.(2013)Wang, Yip, Han, Fok, Lin, Hou, Dong, Hung, Chan, and
  Ho]{0957-4484-24-37-375202}
Wang,~F.; Yip,~S.; Han,~N.; Fok,~K.; Lin,~H.; Hou,~J.~J.; Dong,~G.; Hung,~T.;
  Chan,~K.~S.; Ho,~J.~C. \emph{Nanotechnology} \textbf{2013}, \emph{24},
  375202\relax
\mciteBstWouldAddEndPuncttrue
\mciteSetBstMidEndSepPunct{\mcitedefaultmidpunct}
{\mcitedefaultendpunct}{\mcitedefaultseppunct}\relax
\EndOfBibitem
\bibitem[Parkinson et~al.(2009)Parkinson, Joyce, Gao, Tan, Zhang, Zou,
  Jagadish, Herz, and Johnston]{EnhancementMobilityStructure}
Parkinson,~P.; Joyce,~H.~J.; Gao,~Q.; Tan,~H.~H.; Zhang,~X.; Zou,~J.;
  Jagadish,~C.; Herz,~L.~M.; Johnston,~M.~B. \emph{Nano Lett.} \textbf{2009},
  \emph{9}, 3349--3353\relax
\mciteBstWouldAddEndPuncttrue
\mciteSetBstMidEndSepPunct{\mcitedefaultmidpunct}
{\mcitedefaultendpunct}{\mcitedefaultseppunct}\relax
\EndOfBibitem
\bibitem[Shimamura et~al.(2013)Shimamura, Yuan, Shimojo, and
  Nakano]{EnhancementMobilityStructure02}
Shimamura,~K.; Yuan,~Z.; Shimojo,~F.; Nakano,~A. \emph{Appl. Phys. Lett.}
  \textbf{2013}, \emph{103}, 022105\relax
\mciteBstWouldAddEndPuncttrue
\mciteSetBstMidEndSepPunct{\mcitedefaultmidpunct}
{\mcitedefaultendpunct}{\mcitedefaultseppunct}\relax
\EndOfBibitem
\end{mcitethebibliography}

\end{document}